\documentclass[aps,prl,superscriptaddress,sortedaddress,floatfix,showpacs,reprint]{revtex4-1}

\usepackage{amssymb,amsmath,amsfonts}
\usepackage{mathtools}
\usepackage{mathrsfs}
\usepackage{bbm}
\usepackage{slashed}
\usepackage{nicefrac}

\usepackage{graphicx}
\usepackage[dvipsnames]{xcolor}
\usepackage{array}

\usepackage{simplewick}

\usepackage{hyperref}
\usepackage{xparse}
\usepackage{xspace}

\usepackage{tikz}
\usetikzlibrary{decorations.pathmorphing}
\usetikzlibrary{automata,positioning}

\usepackage{cancel}
\usepackage[normalem]{ulem}

\usepackage{xifthen}
\usepackage{dsfont}
\usepackage[titletoc]{appendix}
\usepackage{booktabs}
\usepackage{units}

\newcommand{\gettitle}{}
\hypersetup{linkcolor=black
	colorlinks,
	linkcolor={red!75!black},
	citecolor={blue!75!black},
	urlcolor={blue!75!black},
	pdftitle={\gettitle},
	pdfauthor={Boussarie, Hatta, Szymanowski, Wallon},
	pdfkeywords={Perturbative QCD} {Small-x}{TMD},
	bookmarksopen=true,
	bookmarksopenlevel=2,
	bookmarksnumbered=true
}
\makeatletter
\newcommand\makebig[2]{%
  \@xp\newcommand\@xp*\csname#1\endcsname{\bBigg@{#2}}%
  \@xp\newcommand\@xp*\csname#1l\endcsname{\@xp\mathopen\csname#1\endcsname}%
  \@xp\newcommand\@xp*\csname#1r\endcsname{\@xp\mathclose\csname#1\endcsname}%
}
\makeatother
\makebig{biggg} {1.3}

\begin{document}

\title{Probing the gluon Sivers function with an unpolarized target: \\ GTMD distributions and the Odderons}% Force line breaks with \\

\author{Renaud Boussarie}
\affiliation{Physics Department, Brookhaven National Laboratory, Upton, NY 11973, USA}

\author{Yoshitaka Hatta}
\affiliation{Physics Department, Brookhaven National Laboratory, Upton, NY 11973, USA}

\author{Lech Szymanowski}
\affiliation{National Centre for Nuclear Research (NCBJ), Warsaw, Poland}

\author{Samuel Wallon}
\affiliation{Laboratoire de Physique Th\'eorique (UMR 8627), CNRS, Univ. Paris-Sud,
Universit\'e Paris-Saclay, 91405 Orsay Cedex, France}
\affiliation{Sorbonne Universit\'{e}, Facult\'e de Physique, 4 place Jussieu, 75252 Paris Cedex 05, France}

\date{\today}% It is always \today, today,
\begin{abstract}
\noindent
It is commonly believed that the Sivers function has uniquely to do with processes involving a transversely polarized nucleon. In this paper we show that this is not necessarily the case. We demonstrate that exclusive pion production in {\it un}polarized electron-proton scattering in the forward region is a direct probe of the gluon Sivers function due to its connection to the QCD Odderon. 
\end{abstract}


\date{\today}
\maketitle
\flushbottom

\paragraph{1. Introduction}

\label{sec:Introduction}

Recently, there has been renewed interest in the interplay between
spin physics and small-$x$ physics, largely motivated by the necessity
to understand the small-$x$ behavior of the helicity distributions
inside a longitudinally polarized proton. Generally speaking, polarization
effects are suppressed by an inverse power of energy compared to unpolarized
ones, but they are enhanced by double logarithms in energy. The resummation
of these logarithms is a particularly challenging problem which has
been first addressed by Bartels, Ermolaev and Ryskin \cite{Bartels:1996wc}.
Recent activities include calculations of various higher order corrections
\cite{Ermolaev:2003zx} and the generalization to the orbital angular momentum
sector \cite{Boussarie:2019icw}, as well as alternative approaches
to resummation \cite{Kovchegov:2015pbl,Kovchegov:2019rrz,Cougoulic:2019aja}.

Meanwhile, another very interesting, and rather unexpected interplay
between spin and small-$x$ physics has been elucidated by Zhou \cite{Zhou:2013gsa}
(see also \cite{Szymanowski:2016mbq,Boer:2015pni,Dong:2018wsp}) in
the transverse polarization sector. It was observed that the gluon Sivers function is related to the Odderon in small-$x$  processes that are sensitive to transverse polarizations. The Sivers function \cite{Sivers:1989cc}
is a transverse momentum dependent (TMD) parton distribution function
that has been exclusively discussed in the context of transverse single
spin asymmetry (SSA). It turns out that the gluonic version of the
Sivers function, despite its obvious connection to spin %polarization,
is not power-suppressed by energy, but instead evolves with leading
single (not double) logarithms at small-$x$. Moreover, this evolution
equation is identical to that for the Odderon amplitude in QCD which
has been well established in the literature \cite{Kovchegov:2003dm,Hatta:2005as,Grabovsky:2013mba,Gerasimov:2012bj}.
The numerical solution of the evolution equation \cite{Yao:2018vcg}
carries various phenomenological implications that can be tested in
future measurements of SSA involving heavy quarks.

In this paper, we dig into this correspondence to a deeper level and
establish relations between the Odderon and the so-called generalized
TMDs (GTMDs) \cite{Meissner:2009ww}. A similar study has been done in Ref.~\cite{Boer:2018vdi}, but the target polarization effect was neglected. 
 We show that, for spin-1/2 hadrons such as the proton, there are three independent
GTMDs that can be associated with the Odderon. Only one of them survives
in the TMD (forward) limit and becomes the gluon Sivers function.
Moreover, we observe that, once reinterpreted as a GTMD, the Sivers
function is not necessarily tied to the transverse polarization of
the incoming state. This opens up an intriguing possibility that one
can experimentally access the gluon Sivers function in \textit{un}polarized
collisions.
Specifically, we calculate the differential cross section $d\sigma/dt$
for the exclusive process $ep\to e'\gamma^{*}p\to e'\pi^{0}p'$ where
both the incoming electron and proton are unpolarized. Since the (virtual)
photon and the pion have opposite $C$-parities, at high energy this
process is dominated by the Odderon exchange which is $C$-odd. We then
show that, among the three Odderon GTMDs that contribute to this process,
the one corresponding to the gluon Sivers function gives the leading
contribution in the forward region $t\approx0$.

This paper is organized as follows. In Section 2, we give a general
discussion on the relation between the Odderon and GTMDs. We show
that there are three types of Odderon GTMDs characterized by different nucleon spinor bilinears. In Section 3, we compute
the cross section of the process $ep\to e' \pi^{0}p'$ and show that it
is dominated by the gluon Sivers function in the forward region. In Section
4, we discuss the properties of the cross section and its measurability in experiments. We finally conclude in Section 5.

\paragraph{2. Odderon and GTMDs \label{sec:equivalence}}

In this section, we clarify the relation between the Odderon amplitude
and gluon GTMD distributions. The general parametrization of dipole-type
gluon GTMDs can be easily adapted from the quark case~\cite{Meissner:2009ww}
and reads~\cite{Bhattacharya:2018lgm}
\begin{eqnarray}
 & & \frac{2M}{\bar{P}^{+}} \int \frac{d^{4}v}{(2\pi)^{3}}\delta(v^{+})e^{ix\bar{P}^{+}v^{-}-i\left(\boldsymbol{k}\cdot\boldsymbol{v}\right)} \nonumber \\ & & \times  \langle P^{\prime}S^{\prime}|{\rm Tr}\left[F^{i+}(-\frac{v}{2})\mathcal{U}_{\frac{v}{2},-\frac{v}{2}}^{\left[+\right]}F^{i+}(\frac{v}{2})\mathcal{U}_{-\frac{v}{2},\frac{v}{2}}^{\left[-\right]}\right]|PS\rangle\label{eq:GTMD}\\
 & & =\bar{u}_{P^{\prime}}[F_{1,1}^{g}+i\frac{\sigma^{i+}}{\bar{P}^{+}}(\boldsymbol{k}^{i}F_{1,2}^{g}+\boldsymbol{\Delta}^{i}F_{1,3}^{g})+i\frac{\sigma^{ij}\boldsymbol{k}^{i}\boldsymbol{\Delta}^{j}}{M^{2}}F_{1,4}^{g}]u_{P},\nonumber 
\end{eqnarray}
where the trace is in the fundamental representation and we defined
the staple gauge links 
\begin{equation}
\mathcal{U}_{x,y}^{\left[\pm\right]}=\left[x^{-},\pm\infty\right]_{\boldsymbol{x}}\left[\boldsymbol{x},\boldsymbol{y}\right]_{\pm\infty}\left[\pm\infty,y^{-}\right]_{\boldsymbol{y}}.\label{eq:staple}
\end{equation}
The notation $[x,y]_{z}$ denotes a straight Wilson line from $y$
to $x$ at fixed $z$. We defined $\bar{P}^{\mu}=\frac{P^{\mu}+P'^{\mu}}{2}$,
$\Delta^{\mu}=P'^{\mu}-P^{\mu}$ is the momentum transfer, and $M$ is the proton mass. The spinors with momenta $P$ and $P^\prime$ have associated spin vectors $S$ and $S^\prime$, which we will write explicitely hereafter. Boldface letters
denote transverse vectors and $i,j=1,2$ are their indices. The four
complex-valued GTMDs $F_{1,n}^{g}$ $(n=1,2,3,4)$ are functions
of $\left(x,\xi,\boldsymbol{k},\boldsymbol{\Delta}\right)$ where
$\xi=-\frac{\Delta^{+}}{2\bar{P}^{+}}$ is the skewness parameter.
The following symmetry properties
are known~\cite{Meissner:2009ww} 
\begin{equation}
F_{1,n}^{g\ast}(x,\xi,\boldsymbol{k}^{2},\boldsymbol{k}\cdot\boldsymbol{\Delta},\boldsymbol{\Delta}^{2})=\pm F_{1,n}^{g}(x,-\xi,\boldsymbol{k}^{2},-\boldsymbol{k}\cdot\boldsymbol{\Delta},\boldsymbol{\Delta}^{2}),\label{eq:f1nstar}
\end{equation}
with a $+$ sign for $n=1,3,4$ and a $-$ sign for $n=2$. In the
following, we shall always be interested in the $\xi\simeq0$ limit
which is typically satisfied in small-$x$ kinematics. %compatible with the eikonal limit. 
 We can then write 
\begin{equation}
F_{1,n}^{g}=f_{1,n}+i\frac{\left(\boldsymbol{k}\cdot\boldsymbol{\Delta}\right)}{M^{2}}g_{1,n},\label{eq:f1n}
\end{equation}
for $n=1,3,4,$ and 
\begin{equation}
F_{1,2}^{g}=\frac{\left(\boldsymbol{k}\cdot\boldsymbol{\Delta}\right)}{M^{2}}f_{1,2}+ig_{1,2},\label{eq:f12}
\end{equation}
where $f$ and $g$ functions are real functions which depend on $x,\boldsymbol{k}^{2},|\boldsymbol{k}\cdot\boldsymbol{\Delta}|,\boldsymbol{\Delta}^{2}$.
In the small $x$ approximation, we will set $x\simeq0$ in the phase
of the l.h.s. of Eq.~(\ref{eq:GTMD}). 
Following the same manipulations as in  \cite{Hatta:2016dxp}, we can rewrite  Eq.~(\ref{eq:GTMD}) as 
\begin{eqnarray}
 & & \int d^{2}\boldsymbol{v}e^{-i\left(\boldsymbol{k}\cdot\boldsymbol{v}\right)}\langle P^{\prime}S^{\prime}|\frac{1}{N_{c}}{\rm Tr}\left[U_{\frac{\boldsymbol{v}}{2}}U_{-\frac{\boldsymbol{v}}{2}}^{\dagger}\right]|PS\rangle \nonumber \\
 & & =(2\pi)^{4}\delta(P^{+}-P'^{+})\frac{\bar{P}^{+}}{2M} \frac{g_s^2}{N_c\left(\boldsymbol{k}^{2}-\frac{\boldsymbol{\Delta}^{2}}{4}\right)} \label{eq:equivfin} \\  & & \times \bar{u}_{P^{\prime}S^{\prime}}\left[F_{1,1}^{g}+i\frac{\sigma^{i+}}{\bar{P}^{+}}(\boldsymbol{k}^{i}F_{1,2}^{g}+\boldsymbol{\Delta}^{i}F_{1,3}^{g})%+i\frac{\sigma^{ij}\boldsymbol{k}^{i}\boldsymbol{\Delta}^{j}}{M^{2}}F_{1,4}^{g}
\right]u_{PS},\nonumber 
\end{eqnarray}
where $U_{\boldsymbol{v}}=[-\infty,\infty]_{\boldsymbol{v}}$. 
Note that the $F_{1,4}^{g}$ term has been neglected in Eq.~(\ref{eq:equivfin}).
Indeed in the eikonal approximation, this term vanishes due to its
$PT$ symmetry \cite{Hatta:2016aoc}. 
The real part of the dipole operator is identified with the Pomeron 
\begin{equation}
\mathcal{P}(\boldsymbol{v})=\mathrm{Re}[\frac{1}{N_{c}}{\rm Tr}(U_{\frac{\boldsymbol{v}}{2}}U_{-\frac{\boldsymbol{v}}{2}}^{\dagger})-1],\label{eq:defpom}
\end{equation}
while the Odderon is the imaginary part
\cite{Hatta:2005as}:
\begin{equation}
{\cal O}(\boldsymbol{v})=i{\rm Im}\frac{1}{N_{c}}{\rm Tr}(U_{\frac{\boldsymbol{v}}{2}}U_{-\frac{\boldsymbol{v}}{2}}^{\dagger}).\label{eq:defodd}
\end{equation}

It is easy to see that, on the l.h.s. of Eq.~(\ref{eq:equivfin}), 
the Pomeron
term is even in $\boldsymbol{k}\leftrightarrow-\boldsymbol{k},$ while
the Odderon term is odd under this exchange.
The latter reads, with the parametrizations from Eqs.~(\ref{eq:f1n},\ref{eq:f12}),
\begin{align}
 & \int d^{2}\boldsymbol{v}e^{-i\left(\boldsymbol{k}\cdot\boldsymbol{v}\right)}\langle P^{\prime},S^{\prime}|{\cal O}(\boldsymbol{v})|P,S\rangle\nonumber \\
 & =g_{s}^{2}(2\pi)^{4}\delta(P^{\prime+}-P^{+})\frac{\bar{P}^{+}}{2M}\frac{\boldsymbol{k}^j}{N_c\left(\boldsymbol{k}^{2}-\frac{\boldsymbol{\Delta}^{2}}{4}\right)}\label{eq:pureodd}\\
 & \times\bar{u}_{P^{\prime},S^{\prime}}\!\left[i\frac{\boldsymbol{\Delta}^j}{M^{2}}g_{1,1}-\frac{\sigma^{i+}}{\bar{P}^{+}}\!\left(\!\delta^{ij}g_{1,2}+\frac{\boldsymbol{\Delta}^{i}\boldsymbol{\Delta}^j}{M^{2}}g_{1,3}\!\right)\right]\!u_{P,S}.\nonumber 
\end{align}
This is the most general parametrization of the dipole-type Odderon
coupling to a generic spin-1/2 hadron. We see that the Odderon is
characterized by three independent GTMDs $g_{1,n}$ ($n=1,2,3$).
It is interesting to notice that by using the Gordon identity in the
first term 
\begin{equation}
\bar{u}_{P^{\prime}S^{\prime}}u_{PS}=\frac{M}{\bar{P}^{+}}\bar{u}_{P^{\prime}S^{\prime}}\gamma^{+}u_{PS}+\bar{u}_{P^{\prime}S^{\prime}}i\sigma^{+i}\frac{\Delta^{i}}{2\bar{P}^{+}}u_{PS},\label{eq:Gordon}
\end{equation}
we recognize a vector coupling between the Odderon and the hadron,
or `vector Odderon' \cite{Ewerz:2013kda}. For the Odderon to be a
pure vector Odderon, we would need the relations $g_{1,1}=2 g_{1,3},\quad g_{1,2}=0$.
However, in general they are independent distributions. We also see that the vector odderon vanishes in the forward limit $\boldsymbol{\Delta}\to 0$.  In the near-forward,
spin non-flip scattering of a proton, the three terms in (\ref{eq:pureodd})
behave like 
\begin{equation}
i(\boldsymbol{k}\cdot\boldsymbol{\Delta})g_{1,1},\quad(\boldsymbol{k}\times\boldsymbol{S})_{z}g_{1,2},\quad(\boldsymbol{k}\cdot\boldsymbol{\Delta})(\boldsymbol{\Delta}\times\boldsymbol{S})_{z}g_{1,3}.\label{po}
\end{equation}
Since the Odderon amplitude is a scalar function and is odd under
$\boldsymbol{k}\to-\boldsymbol{k}$, it should be proportional to
the vector $\boldsymbol{k}$, contracted by another two-dimensional
vector. (\ref{po}) exhausts all possibilities in this kinematics \footnote{Note that we neglected the $F_{1,4}$ due to the eikonal limit, otherwise the $(\boldsymbol{k}\times\boldsymbol{\Delta})_{z}$ contraction would also appear in this list}.
$g_{1,1}$, to which we will refer as the vector Odderon term, is
associated with momentum transfer $\boldsymbol{\Delta}$ (or impact
parameter $\boldsymbol{b}$ in the coordinate space). This exists
even for spinless hadrons, and the Odderon coupling discussed in most
literature \cite{Czyzewski:1996bv,Kovchegov:2003dm,Kovchegov:2012ga,Zhou:2013gsa,Boer:2018vdi,Dumitru:2018vpr,Dumitru:2019qec}
is of this type. $g_{1,2}$ is the spin-dependent Odderon \cite{Zhou:2013gsa}.
%sometimes referred to as the spin Odderon. 
To our knowledge, the structure
of the last term $g_{1,3}$, let us call it the spin-vector Odderon,
has not been identified in the literature. This features a sine correlation
between the azimuthal angles of $\boldsymbol{\Delta}$ and $\boldsymbol{S}$
which may lead to distinct experimental signals.

In the forward $\boldsymbol{\Delta}=\boldsymbol{0}$ limit, only $g_{1,2}=\mathrm{Im}(F_{1,2}^{g})$
survives. If we further assume that the proton is transversely polarized, we
find the familiar structure of the Sivers function 
\begin{equation}
\boldsymbol{k}^{2}\!\int\!\! d^{2}\boldsymbol{v}e^{-i\left(\boldsymbol{k}\cdot\boldsymbol{v}\right)}\langle PS|{\cal O}(\boldsymbol{v})|PS\rangle\propto2\frac{(\boldsymbol{k}\times\boldsymbol{S})_{z}}{M^{2}}\mathrm{Im}(F_{1,2}^{g}),\label{eq:forwardodderon}
\end{equation}
with the identification 
\begin{equation}
2\mathrm{Im}(F_{1,2}^{g})(x,0,\boldsymbol{k},\boldsymbol{0})=-xf_{1T}^{\perp g}(x,\boldsymbol{k}^{2}).\label{eq:TMDlimit}
\end{equation}
Interestingly, and importantly for our purpose, the relevant spinor
product is nonvanishing also for a generic helicity flip contribution
\begin{equation}
\boldsymbol{k}^{i}\bar{u}_{PS'}\sigma^{+i}u_{PS}\propto(\boldsymbol{k}\times\boldsymbol{S})_{z}\delta_{h,-h^{\prime}},\label{eq:usigmau}
\end{equation}
where $\boldsymbol{S}=\frac{1}{\sqrt{2}}\left(1,ih\right)$ is a transverse
vector built from proton helicity $h=\pm1$ which resembles a circular
polarization vector in the transverse plane. We thus find that the
gluon Sivers function appears not only in processes involving a transversely
polarized proton, but also in processes where the proton spin flips.
This includes unpolarized collisions where one independently sums
over the polarizations of the initial and final proton helicities.
This is the key observation that will be further explored in the next
sections.

\paragraph{3, Exclusive diffractive pion production $ep\to e'\pi^{0}p'$ at small $x$\label{sec:DVMP}}

\begin{figure}
\includegraphics[width=0.31\textwidth]{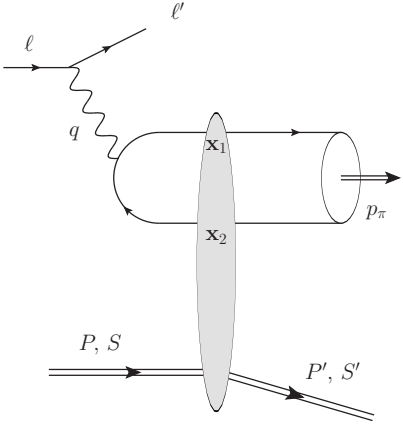}\caption{Electroproduction of a $\pi^{0}$ meson at small $x$. The gray blob represents the interaction with the background field, and the white
blob represents the Distribution Amplitude of the meson.\label{fig:diagram}}
\end{figure}

In this section, we compute the differential cross section of exclusive
pion production $ep\to e'\pi^{0}p'$ at high energy. The process is
depicted in Fig.~\ref{fig:diagram}. Since the virtual photon and
the pion have negative and positive $C$-parity, respectively, the exchanged
object in the $t$-channel has negative $C$-parity. The process is thus
a classic example of possible Odderon signatures \cite{Berger:1999ca}.
Instead of $\pi^{0}$, one can also consider $\eta_{c}$ \cite{Czyzewski:1996bv}.
However, the rate will be smaller and $\eta_{c}$ is more difficult
to reconstruct. 
We will use the standard variables for Deeply Virtual Meson Production (DVMP)%, in terms of the kinematics of Figure~\ref{fig:diagram}:
\begin{equation}
Q^{2}=-q^{2},\quad x_{B}=\frac{Q^{2}}{2\left(P\cdot q\right)},\quad t=\Delta^{2},\quad y=\frac{\left(P\cdot q\right)}{\left(P\cdot\ell\right)}. \label{variables}
\end{equation}
Not that in the eikonal approximation, $t\approx -\boldsymbol{\Delta}^2$.
The computation of the amplitude at small $x$ performed in this article
relies on the covariant background field methods developed in~\cite{Balitsky:1995ub,Balitsky:1998kc,Balitsky:1998ya}, very similar to the Color Glass Condensate (CGC) approach~\cite{McLerran:1993ni,McLerran:1993ka,McLerran:1994vd}.  The outgoing quark and antiquark fields in the shock wave background created by the target gluon fields are 
\begin{eqnarray}
& & \bar{\psi}_{eff}(x_{0}) = \theta(x_{0}^{-})\bar{\psi}(x_{0}) \label{eq:effquark} \\ & & + \theta(-x_{0}^{-})\int d^{4}x_{1}\delta(x_{1}^{-})\bar{\psi}(x_{1})\gamma^{-}G(x_{10})(U_{\boldsymbol{x}_{1}}-1) \nonumber
\end{eqnarray}
for the quark, and 
\begin{eqnarray}
& & \psi_{eff}(x_{0}) =\theta(x_{0}^{-})\psi(x_{0}) \label{eq:effantiquark} \\ & & -\theta(-x_{0}^{-})\int d^{4}x_{2}\delta(x_{2}^{-})G(x_{02})\gamma^{-}\psi(x_{2})(U_{\boldsymbol{x}_{2}}^{\dagger}-1) \nonumber
\end{eqnarray}
for the antiquark, where $x_{1,2}^-=0$ is the time of interaction with the shockwave and $G$ is the free fermion propagator. From these two effective fields, we can build the
amplitude for DVMP as 
\begin{eqnarray}
\mathcal{A} & = & \int d^{4}x_{0}\bar{u}_{\ell^{\prime}}(-ie_{\ell})\gamma^{\mu}u_{\ell}G_{\mu\nu}(q)e^{-i\left(q\cdot x_{0}\right)} \nonumber \\ & \times & \langle P^{\prime}\pi^0|\bar{\psi}_{eff}(x_{0})(-ie_{f})\gamma^{\nu}\psi_{eff}(x_{0})|P\rangle,\label{eq:ampini}
\end{eqnarray}
where $e_f$ is an effective electric charge which takes into account the flavor content of the meson. With the explicit expressions from Eqs.~(\ref{eq:effquark}, \ref{eq:effantiquark})
and with a Fierz decomposition in color space and in spinor space
on a complete set of $\Gamma^{\lambda}$ matrices, Eq.~(\ref{eq:ampini}) becomes: 
\begin{align}
\mathcal{A} & =\frac{e_{f}e_{\ell}}{4N_{c}}\bar{u}_{\ell^{\prime}}\gamma^{\mu}u_{\ell}\int d^{4}x_{0}d^{4}x_{1}d^{4}x_{2}\theta(-x_{0}^{-})\delta(x_{1}^{-})\delta(x_{2}^{-})\nonumber \\
 & \times\langle\pi^0|\bar{\psi}(x_{1})\Gamma^{\lambda}\psi(x_{2})|0\rangle\langle P^{\prime}|\mathrm{Tr}(U_{\boldsymbol{x}_{1}}U_{\boldsymbol{x}_{2}}^{\dagger})-N_{c}|P\rangle\label{eq:amp2}\\
 & \times e^{-i\left(q\cdot x_{0}\right)} G_{\mu\nu}(q)\mathrm{Tr}\left[\gamma^{-}G(x_{10})\gamma^{\nu}G(x_{02})\gamma^{-}\Gamma_{\lambda}\right].\nonumber 
\end{align}
We evaluate this at leading $s$-channel twist by taking the light cone expansion
of the bilocal vacuum-to-meson correlator $\left\langle \pi^0\left|\bar{\psi}\psi\right|0\right\rangle $.
Projecting onto the axial-vector state $(\Gamma^{\lambda})(\Gamma_{\lambda})\rightarrow(\gamma^{\lambda}\gamma_{5})(\gamma_{5}\gamma_{\lambda})$ to take into account the properties of the leading twist chiral-even DA for the pion
\begin{align}
& \langle\pi(p_{\pi})|\bar{\psi}(x_{1})\gamma^{\lambda}\gamma_{5}\psi(x_{2})|0\rangle \label{eq:DA} \\
& =if_{\pi}p_{\pi}^{\lambda}\int_{0}^{1}dze^{iz(p_{\pi}\cdot
x_{1})+i\bar{z}(p_{\pi}\cdot x_{2})}\phi_{\pi}(z), \nonumber
\end{align}
and following computation steps very similar to what is described in~\cite{Boussarie:2018zwg}, one gets: 
\begin{align}
\mathcal{A} & =\frac{-e_{f}e_{\ell}f_{\pi}}{N_{c}}\delta(q^{-}\!-p_{\pi}^{-})\bar{u}_{\ell^{\prime}}\gamma^{\mu}u_{\ell}\!\int \!\!d^{2}\boldsymbol{b}e^{i(\boldsymbol{q}-\boldsymbol{p}_{\pi})\cdot\boldsymbol{b}}\nonumber \\
 & \times \epsilon^{\alpha\beta+-} \int\!\! d^{2}\boldsymbol{r}\!\int_{0}^{1}\!dz\phi_{\pi}(z)r_{\perp\alpha}\sqrt{\frac{z\bar{z}Q^{2}}{\boldsymbol{r}^{2}}}K_{1}(\sqrt{z\bar{z}Q^{2}\boldsymbol{r}^{2}})\nonumber\\
 & \times\langle P^{\prime}|\mathrm{Tr}(U_{\boldsymbol{b}+\bar{z}\boldsymbol{r}}U_{\boldsymbol{b}-z\boldsymbol{r}}^{\dagger}\!-\!1)|P\rangle G_{\mu\nu}(q)\!\left(q^{-}g_{\perp\beta}^{\nu}-n^{\nu}q_{\perp\beta}\right)\!,\label{eq:amp6} 
\end{align}
where $\bar{z}=1-z$. $\phi_\pi$ is the pion's leading twist axial-vector distribution amplitude (DA) and $f_\pi=131$ MeV is the decay constant. 
As expected, the amplitude~(\ref{eq:amp6}) involves the $t$ channel
exchange of an Odderon. Indeed, the symmetry property of the 
DA $ \phi_{\pi}(z)=\phi_{\pi}(\bar{z}) $ means that the $\boldsymbol{r}\leftrightarrow-\boldsymbol{r}$ antisymmetric
contribution from the dipole matrix element contributes, i.e. its
imaginary part is selected by the $C$ parity of the process. To make the connection to the Odderon GTMDs  explicit, we use the formula 
\begin{eqnarray}
 & & \langle P^{\prime}|\mathrm{Tr}(U_{\boldsymbol{b}+\bar{x}\boldsymbol{r}}U_{\boldsymbol{b}-x\boldsymbol{r}}^{\dagger})-N_{c}|P\rangle \nonumber \\
 & = & \int\!\frac{d^{2}\boldsymbol{k}}{\boldsymbol{k}^{2}-\frac{\boldsymbol{\Delta}^{2}}{4}}\int\!\frac{d^{2}\boldsymbol{v}}{(2\pi)^{2}}e^{-i\left(\boldsymbol{k}\cdot\boldsymbol{v}\right)}e^{i\left(\boldsymbol{k}+\frac{x-\bar{x}}{2}\boldsymbol{\Delta}\right)\cdot\boldsymbol{r}} \nonumber \\ 
 & \times & e^{-i\left(\boldsymbol{\Delta}\cdot\boldsymbol{b}\right)} \langle P^{\prime}|\mathrm{Tr}(\partial^{i}U_{\frac{\boldsymbol{v}}{2}})(\partial^{i}U_{-\frac{\boldsymbol{v}}{2}})|P\rangle,\label{eq:Equiv-1}
\end{eqnarray}
which is a simple case of the general formula derived in~\cite{Altinoluk:2019wyu} and is compatible with the results of Section 2. Using these results, Eq.~(\ref{eq:amp6}) then becomes 
\begin{align}
\mathcal{A} & =\frac{e_{f}e_{\ell}f_{\pi}}{N_{c}}g_{s}^{2}(2\pi)^{5}\delta(p_{\pi}+\Delta-q)\bar{u}_{\ell^{\prime}}\gamma^{\mu}u_{\ell}\nonumber \\
 & \times G_{\mu\nu}(q)\epsilon^{\alpha\beta+-}\left(q^{-}g_{\perp\beta}^{\nu}-n^{\nu}q_{\perp\beta}\right) \int_{0}^{1}dz\phi_{\pi}(z) \nonumber \\
 & \times\frac{\bar{P}^{+}}{2M}\int\frac{d^{2}\boldsymbol{k}}{\boldsymbol{k}^{2}-\frac{\boldsymbol{\Delta}^{2}}{4}}\frac{\boldsymbol{k}^j\left(k_{\perp\alpha}+\frac{z-\bar{z}}{2}\Delta_{\perp\alpha}\right)}{\left(\boldsymbol{k}+\frac{z-\bar{z}}{2}\boldsymbol{\Delta}\right)^{2}+z\bar{z}Q^{2}}\label{eq:ampGTMD}\\
 & \times\bar{u}_{P^{\prime},S^{\prime}}\!\left[i\frac{\boldsymbol{\Delta}^j}{M^{2}}g_{1,1}-\frac{\sigma^{i+}}{\bar{P}^{+}}\!\left(\!\delta^{ij}g_{1,2}+\frac{\boldsymbol{\Delta}^{i}\boldsymbol{\Delta}^j}{M^{2}}g_{1,3}\!\right)\right]\!u_{P,S}.\nonumber 
\end{align}
Let us focus on the small $\boldsymbol{\Delta}$ limit in which Eq.~(\ref{eq:ampGTMD}) simplifies drastically. Only $g_{1,2}\propto f_{1T}^{\perp g}$ survives in this limit and the amplitude becomes proportional to % and  reduces to 
\begin{eqnarray}
\mathcal{A}&\propto& \int\frac{d^{2}\boldsymbol{k}}{\boldsymbol{k}^{2}+z\bar{z}Q^{2}}xf_{1T}^{\perp g}(x,\boldsymbol{k}^{2})\label{eq:ampfor}\\
&=& -\frac{1}{z\bar{z}Q^2}\int  \frac{d^2\boldsymbol{k} \,\boldsymbol{k}^{2}}{\boldsymbol{k}^{2}+z\bar{z}Q^{2}} xf_{1T}^{\perp g}(x,\boldsymbol{k}^{2}),
\nonumber 
\end{eqnarray}
where in the second line we used the relation 
\begin{equation}
\int d^2\boldsymbol{k} \, xf_{1T}^{\perp g}(x,\boldsymbol{k}^2)=0. \label{ji}
\end{equation}
This relation was first noted in \cite{Zhou:2013gsa} within a specific model, but it it is actually a general result \footnote{J. Zhou, private communications.}.
The factor of $\boldsymbol{k}^2$ in the numerator ensures gauge invariance of the amplitude. 

In order to get the cross section, one has to square  Eq.~(\ref{eq:ampGTMD}) and sum over the initial and final proton spins. As can be seen  from Eq.~(\ref{eq:usigmau}),  a nonvanishing contribution arises when the proton flips spin $S=-S'$.
Taking into account the quark content of the pion $\left|\pi^0\right\rangle = (\left| u \bar{u} \right\rangle - \left| d\bar{d}\right\rangle)/\sqrt{2}$ so that $e_f = (e_u-e_d)/\sqrt{2}=e/\sqrt{2}$, we thus arrive at the main result of this paper 
\begin{eqnarray}
& & \frac{d\sigma}{dx_B dQ^{2}d\left|t\right|}  =\frac{\pi^5\alpha_{\mathrm{em}}^{2}\alpha_{s}^{2}f_{\pi}^{2}}{2^3x_B N_{c}^{2}M^{2}Q^{6}}(1-y+\frac{y^{2}}{2})\label{eq:XSfin}\\
& & \times\left[\int_{0}^{1}dz\frac{\phi_{\pi}(z)}{z\bar{z}}\int d\boldsymbol{k}^{2}\frac{\boldsymbol{k}^{2}}{\boldsymbol{k}^{2}+z\bar{z}Q^{2}}xf_{1T}^{\perp g}(x,\boldsymbol{k}^{2})\right]^{2}, \nonumber 
\end{eqnarray}
valid in the forward region $t\approx 0$. The corrections to this formula are of order $t/M^2$ where $M$ is the target mass. Quite remarkably, the leading contribution to this {\it un}polarized observable is given by the gluon Sivers function which is usually associated with a transversely polarized nucleon, and this has been missed in all the previous calculations.  
 We expect that a similar conclusion holds in related observables such as $\eta_c$ production. 
Note that only the transversely polarized virtual photon contributes to this formula. The gamma matrix trace associated with the quark loop vanishes for the longitudinally polarized virtual photon  even in  nonforward kinematics as a consequence of the eikonal coupling of the $t$-channel gluons to quarks and the presence of a $\gamma_5$.

In the large-$Q^2$ region, the cross section is directly related to the $C$-odd collinear three-gluon correlator $O(x_1,x_2)$ relevant to single spin asymmetry \cite{Beppu:2010qn,Zhou:2013gsa}
\begin{equation}
\int d\boldsymbol{k}^{2}\boldsymbol{k}^{2}xf_{1T}^{\perp g}(x,\boldsymbol{k}^{2}) \propto O(x,x)+O(x,0).
\end{equation}
Note, however, that neglecting $\boldsymbol{k}^2$ in the denominator of Eq.~(\ref{eq:XSfin}) results in an end-point singularity at $z,\bar{z}=0$. In practice, this should be cutoff at $z,\bar{z}\sim \Lambda^2/Q^2$, leading to a logarithmic enhancement $\ln^2 Q^2/\Lambda^2$ at large $Q^2$. 
 As an example, consider the asymptotic form $\phi_{\pi}(z)=6z\bar{z}$ and a simple model
for the Sivers function at small $x$ built in \cite{Zhou:2013gsa} 
\begin{equation}
xf_{1T}^{\perp g}(x,\boldsymbol{k}^{2})=\mathcal{C}\frac{\boldsymbol{k}^{2}}{\Lambda^4}\left(2-\frac{\boldsymbol{k}^{2}}{\Lambda^{2}}\right)e^{-\boldsymbol{k}^{2}/\Lambda^{2}},\label{eq:fmodel}
\end{equation}
 where $\mathcal{C}$ is a dimensionless constant. We find
\begin{eqnarray}
& & \frac{d\sigma}{dx_B dQ^{2}d\left|t\right|} =\pi^5\frac{2\alpha_{\mathrm{em}}^{2}\alpha_{s}^{2}f_{\pi}^{2}Q^{2}}{x_B M^{2}}(1-y+\frac{y^{2}}{2})\label{eq:XSfin-1}\\
& & \times\frac{\mathcal{C}^{2}}{\Lambda^8}\left[\int d\alpha\frac{\alpha^{2}\left(2-\frac{Q^{2}}{\Lambda^{2}}\alpha\right)}{\sqrt{1+4\alpha}}\ln\left(\frac{\sqrt{1+4\alpha}+1}{\sqrt{1+4\alpha}-1}\right)e^{-\alpha\frac{Q^{2}}{\Lambda^{2}}}\right]^{2}.\nonumber 
\end{eqnarray}
For $\Lambda\ll Q$, the exponent peaks the integrand around $\alpha=0$,
which allows for a fully analytical integration:
\begin{align}
\frac{d\sigma}{dx_B dQ^{2}d\left|t\right|} & =\pi^5\frac{2\alpha_{\mathrm{em}}^{2}\alpha_{s}^{2}f_{\pi}^{2}}{x_B M^{2}}(1-y+\frac{y^{2}}{2})\label{eq:XSfin-1-1}\\
 & \times\mathcal{C}^{2}\frac{\Lambda^{4}}{Q^{10}}\left[5-2\gamma_{E}+2\log\left(\frac{\Lambda^{2}}{Q^{2}}\right)\right]^{2}.\nonumber 
\end{align}

\paragraph{4. Discussions}

 That the cross section $d\sigma/dt$ does not vanish in the forward limit $\boldsymbol{\Delta}\to 0$ is quite  nontrivial. As we mentioned above, only the transversely polarized virtual photon contributes to the cross section. The  polarization vector of the photon must then be contracted with another transverse vector. For the case of pion production, and without considering  proton spin effects, the only available vector is  momentum transfer $\boldsymbol{\Delta}$. This is why the Odderon contribution to $d\sigma/dt$ vanishes in the $t \approx -\boldsymbol{\Delta}^2 \to 0$ limit in Refs.~\cite{Czyzewski:1996bv,Engel:1997cga,Dumitru:2019qec} (see also \cite{Harland-Lang:2018ytk}). In contrast, Ref.~\cite{Berger:1999ca} obtained a finite result in the same limit. This is because the authors of \cite{Berger:1999ca} considered processes where the proton is excited to a negative parity resonance $N^*$ which is modeled as a $p$-wave bound state of a diquark and a quark. The $p$-wave wavefunction involves a transverse vector which can be contracted with the photon polarization vector, thereby giving a finite result at $t=0$. We however note that considering the excitation and decay of the target proton  introduces extra theoretical uncertainties.    

Our central observation in this paper is that the required vector can come from the spin-$1/2$ nature of the proton. The spin-dependent Odderon accompanies the spinor product $\bar{u}\sigma^{+i}u$ which carries a transverse vector index. This is nonvanishing provided the proton flips spin $S'=-S$ (see Eq.~(\ref{eq:usigmau})), and such spin-flip contributions are automatically included when calculating unpolarized cross sections. One may wonder why the proton can flip helicity in the eikonal approximation in the forward limit. Eq.~(\ref{eq:pureodd}) shows that this can occur nonperturbatively, assisted by the transverse vector $\boldsymbol{k}$ which carries one unit of angular momentum. It is also a necessary consequence of the existence of the spin-dependent Odderon.

We should add that our calculation is valid when $Q^2$ is perturbative, say, $Q^2>1$ GeV$^2$ since this is the only hard scale. Some might be tempted to assume that at sufficiently high energy, the gluon saturation scale $Q_s$ could serve as a hard scale. However, as demonstrated in \cite{Yao:2018vcg}, the characteristic momentum scale of the Odderon amplitude does not scale with $Q_s$.

 The previous Odderon search at HERA \cite{Adloff:2002dw} measured neutrons from the reaction $p\to N^* \to n$ based on a calculation in Ref.~\cite{Berger:1999ca}, and no signal was observed. We simply propose to measure the elastically scattered proton in the final state. In the near forward region, one should see the  flattening of the curve $d\sigma/dt$ as $t\to 0$ before reaching the (diverging) contribution from the Primakoff process $\gamma^*\gamma \to \pi^0$ in the small-$t$ limit: 
\begin{equation}
\frac{d\sigma^{Primakoff}}{dx_{B}dQ^{2}d\left|t\right|} \approx \frac{(2\pi)\alpha_{em}^{4}f_{\pi}^{2}(1+(1-y)^{2})}{x_{B}Q^{6}\left|t\right|}\left[\int_{0}^{1}dz\frac{\phi_{\pi}(z)}{\phi_{\pi}^{\infty}(z)}\right]^{2}\label{pri}
\end{equation}
where $\phi_{\pi}^{\infty}$ is the asymptotic DA $\phi_\pi^\infty(z)=6z\bar{z}$. Despite the relative suppression factor $\alpha_{em}^2$, this can be a serious background at small-$t$ and large-$Q^2$. Fortunately, there is no interference between the spin-dependent Odderon and the (leading) Primakoff amplitude because the proton helicity flips in the former but not in the latter. Therefore, one can simply subtract (\ref{pri}) from the measured cross section.  For larger values of $t$, the Primakoff process should be negligible and one can probe  the three types of Odderon GTMDs $g_{1,1}$, $g_{1,2}$ and $g_{1,3}$. However, the cross section formula in this region is rather complicated. 

\paragraph{5. Conclusions}

In this paper, we have shown that GTMDs provide a unified framework to treat the spin-dependent and spin-independent Odderons on an equal footing. The formula (\ref{eq:pureodd})
which involves three independent dipole gluon GTMDs describes how
the Odderon couples to generic spin-1/2 hadrons. While discussions
of Odderon-hadron coupling are scattered in the literature, it has
not been previously presented in this most general, coherent form.
Of particular interest is the spin-dependent Odderon $g_{1,2}$ which
reduces to the gluon Sivers function in the forward (TMD) limit \cite{Zhou:2013gsa}.
We have demonstrated that this function gives the dominant contribution to exclusive pion production
$ep\to e'\pi^{0}p'$ in the forward region $t\approx0$.

 It would be very interesting to test our prediction at existing and future $ep$ colliders such as the EIC, and especially, the LHeC where the target polarization is not planned at the moment.  The detection of events will be an unambiguous signal of the much sought-after QCD Odderon, and at the same time, shed light on the magnitude of the gluon Sivers effect which remains largely mysterious to date.

A lot of progress has been made for one-loop corrections to exclusive diffractive processes at small $x$~\cite{Boussarie:2014lxa,Boussarie:2016ogo,Boussarie:2016bkq,Boussarie:2019ero}. In particular, a similar computation to that of~\cite{Boussarie:2016bkq}, adapted from results of~\cite{Boussarie:2016ogo}, would easily provide the one-loop corrections to the process described in this article, furthering the precision for EIC and LHeC predictions.
%%%%%%%%%%%%%%%

\paragraph*{Acknowledgements}

The authors are grateful to A.V. Grabovsky for his collaboration at the early stages of this work, and to B. Pire for insightful discussions. This work is supported by the U.S. Department of Energy, Office of
Science, Office of Nuclear Physics, under contract No. DE- SC0012704, and in part by Laboratory Directed Research and Development (LDRD) funds from Brookhaven Science Associates. This project has received funding from the European Unions Horizon 2020 research and innovation program STRONG-2020 under grant Agreement No. 824093. L.S. thanks the CNRS for support. The work of L.S. is also partially supported by Grant No. 2017/26/M/ ST2/01074 of the National Science Center in Poland and by French-Polish Collaboration Agreement POLONIUM.

\bibliographystyle{apsrev}
\bibliography{MasterBibtex}

\end{document}